\documentclass{iaus}
\usepackage{graphicx}

\title[S265.~~Detailed analyses of three CEMP stars] %% give here short title %%
{Detailed analyses of three neutron-capture-rich carbon-enhanced
  metal-poor stars}

\author[N.T. Behara, P. Bonifacio, H.-G. Ludwig et al.]   %% give here short author list %%
{N.T.Behara$^{1,2}$, 
       P.Bonifacio$^{1,2,3}$,
       H.-G. Ludwig$^{1,2}$,
       L.Sbordone$^{1,2}$,
       J.I.Gonz\'alez Hern\'andez$^{1,2}$,
       \and E.Caffau$^2$}

\affiliation{$^1$CIFIST Marie Curie Excellence Team \\
        $^2$GEPI, Observatoire de Paris CNRS, Universit\'e Paris Diderot
        $^3$Istituto Nazionale di Astrofisica - Osservatorio Astronomico di
Trieste\\
email: {\tt natalie.behara@obspm.fr}}

\pubyear{2009}
\volume{xxx}  %% insert here IAU Symposium No.
\jname{Title of your IAU Symposium}
\editors{A.C. Editor, B.D. Editor \& C.E. Editor, eds.}
\begin{document}

\maketitle

\begin{abstract}
Approximately 20\% of very metal-poor stars ([Fe/H] $<$ --2.0) are strongly enhanced in
carbon ([C/Fe] $>$ +1.0). Such stars are referred to as carbon-enhanced
metal-poor (CEMP) stars. We present a chemical abundance analysis based on
high resolution spectra acquired with UVES at the VLT of three dwarf CEMP
stars: SDSS J1349-0229, SDSS J0912+0216 and SDSS J1036+1212. These very
metal-poor stars, with [Fe/H] $<$ -2.5, were selected from our ongoing survey
of extremely metal-poor dwarf candidates from the SDSS. 

Among these CEMPs, SDSS J1349-0229 has been identified as a carbon star ([C/O] $>$ +1.0).
First and second peak $s$-process elements, as well as second peak $r$-process elements
have been detected in all stars. In addition, elements from the third
$r$-process peak were detected in one of the stars, SDSS J1036+1212.
We present the abundance results of these stars in the context of
neutron-capture nucleosynthesis theories.

\end{abstract}

\firstsection % if your document starts with a section,
              % remove some space above using this command.
\section{Introduction \& analysis}

\begin{figure}[]
% \vspace*{-2.0 cm}
\begin{center}
 \includegraphics[width=.48\columnwidth]{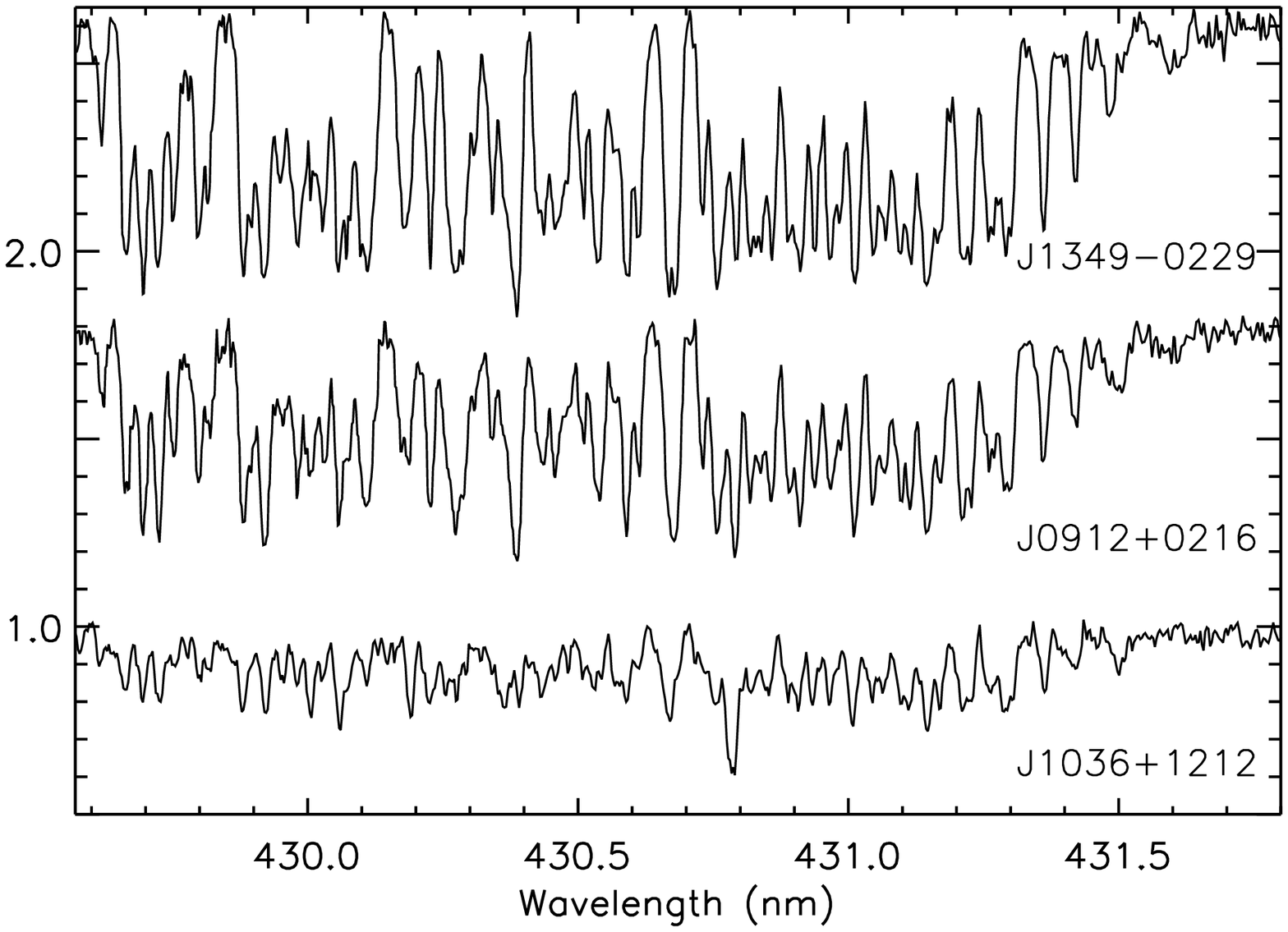} 
 \includegraphics[width=.48\columnwidth]{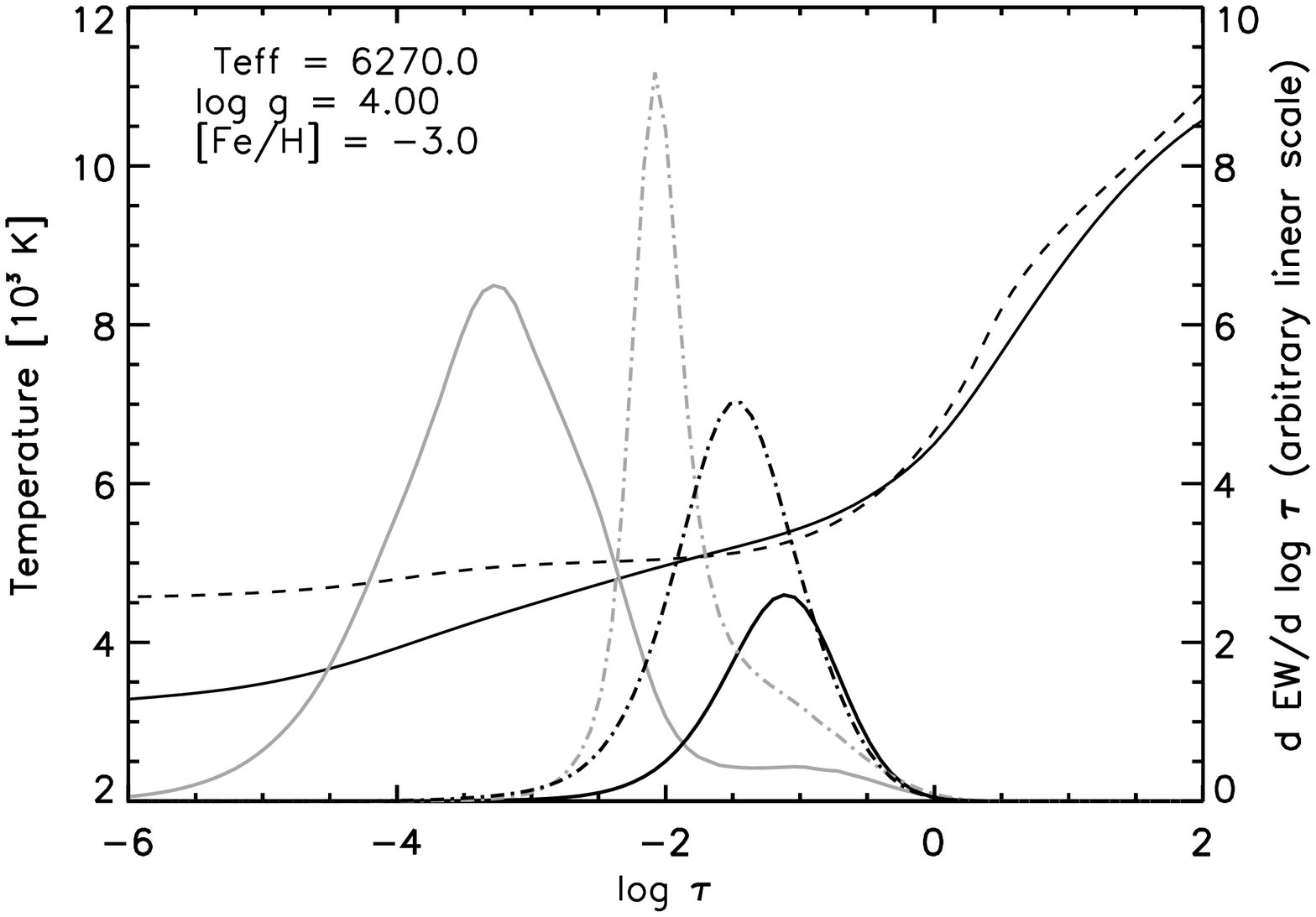} 
 \vspace*{-0.25 cm}
 \caption{{\it Left figure:} Observed spectra of the CH {\it G} band.
{\it Right figure:} Equivalent width contribution function plotted as a
  function of optical depth for the 3D model (grey) and the 1D$_{\rm LHD}$
  model (black) for two different C/O ratios. Scaled solar
  C/O is plotted as a solid line, while a C/O ratio typical for a
  CEMP star is plotted as a dot-dashed line.  
  Overplotted are the average temperature profile of the 3D model (solid
  line) and of the 1D$_{\rm LHD}$ model (dashed line). 
}
   \label{fig1}
\end{center}
\end{figure}

The objects SDSS J1349-0229, SDSS J0912+0216 and SDSS J1036+1212 were selected
as candidates from the Sloan Digital Sky Survey as part of our ongoing survey
of stars at low metallicity. High resolution UVES spectra were obtained which
revealed that these are dwarf CEMP stars with [C/Fe]$\,>
1.0$. Figure~\ref{fig1} displays the CH {\it G} bands  
of all three stars, showing clearly the C enhancement.

The atmospheric parameters were determined using an LTE 1D analysis. 
ATLAS model atmospheres and SYNTHE (Kurucz, 1993)
synthetic spectra have been employed in the analysis. 
Lines of CH, NH and OH were used to determine the carbon, nitrogen and oxygen abundances. We 
employed 3D model atmospheres, computed with the CO$^5$BOLD code (Freytag
et al.~2002; Wedemeyer et al.~2004). The spectral synthesis calculations were
performed with the code Linfor3D. 
Details regarding the 3D molecular calculations and results can be found in
\cite{behara2009}. Other elements were investigated using 1D model atmospheres. 
Adopted stellar parameters and a summary of the abundances measured are listed in Table~\ref{tab1}.  

%\vspace*{-0.25 cm}
\begin{table}
  \begin{center}
  \caption{Adopted stellar parameters and abundances, where [ ] denotes 3D abundances.}
  \label{tab1}
 {\scriptsize
\begin{tabular}[t]{l c c c c c c c c c}
%\hline
Star       & $T_{\rm eff}$ & log $g$ & [Fe/H] & [C/Fe]      & [N/Fe]       & [O/Fe]     & [Sr/Fe] & [Ba/Fe] & [Eu/Fe]\\
\hline
J1349--0229 & 6200        & 4.00     & --3.0  &  2.82 [2.09] & 1.60 [0.67] & 1.88 [1.69] & 1.35 & 2.26 & 1.67\\
J0912+0216 & 6500        & 4.50     & --2.5  &  2.17 [1.67] & 1.75 [1.07] &            & 0.53 & 1.58 & 1.25\\
J1036+1212 & 6000        & 4.00     & --3.2  &  1.47 [0.96] & 1.29 [0.51] &            & --0.51& 1.26 & 1.31\\
%\hline
\end{tabular}
  }
 \end{center}
\vspace{1mm}
\end{table}

\section{Abundances and comparison with similar stars}

\begin{figure}[]
 \vspace*{-0.25 cm}
\begin{center}
 \includegraphics[width=.48\columnwidth]{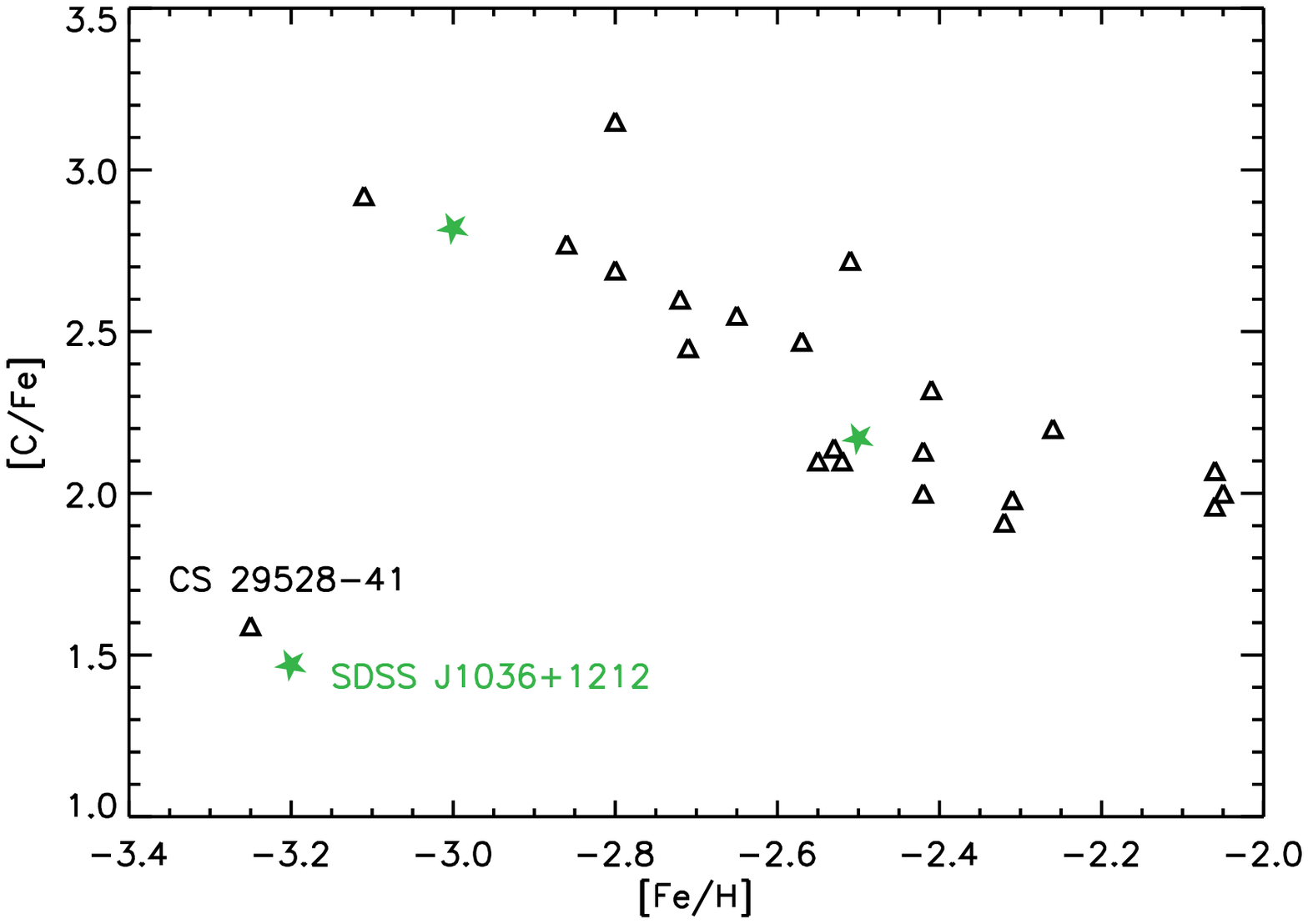} 
 \includegraphics[width=.48\columnwidth]{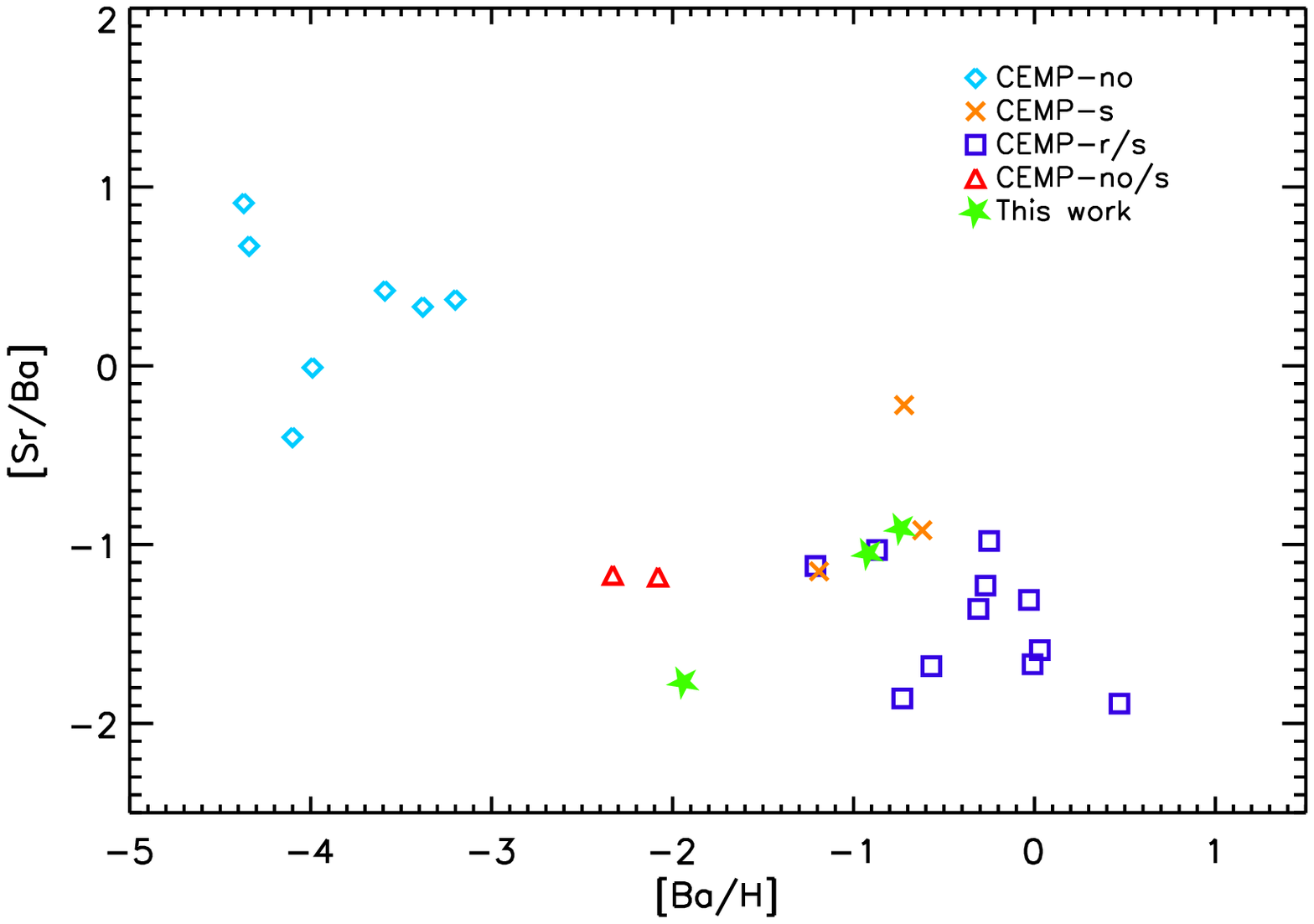} 
 \vspace*{-0.25 cm}
% \vspace*{-1.0 cm}
 \caption{{\it Left figure:} We compared [C/Fe] of our stars (star symbols) to a
   sample of CEMP stars from Sivarani et al.~(2006). Excluding the 
  two most metal-poor stars, a clear correlation is seen between [C/Fe] and
  [Fe/H]. The two exceptions are SDSS J1036+1212 from this work 
  and a CEMP-no/s star CS 29528-041.
  {\it Right figure:} We attempt to classify our stars by comparing their
  [Ba/Sr] abundance against the different families of CEMP stars. We classify SDSS
  J1349-0229 and SDSS J0912+0216 as CEMP-r+s stars due to their high Ba and Eu
(both $>$ 1.0). SDSS J1036+1212 becomes the third member of the CEMP-no/s
class, due to its low Sr abundance.
}
   \label{fig3}
 \vspace*{-0.25 cm}
\end{center}
\end{figure}

The calculated 3D correction for OH is quite small compared to values found in
literature for metal-poor stars (Asplund \& Garcia Perez, 2001). The
corrections for OH are very sensitive to the carbon enhancement in the
atmosphere. In Fig.~\ref{fig3} we plot the contribution 
function for an OH line computed for two different C/O ratios. In the typical CEMP case,
no OH is formed higher in the atmosphere, since due to the high C
content, the oxygen is tied up in CO in this region. 
The stars of this work are presented in the context of the different classes
of CEMP stars in Fig.~\ref{fig3}.

\vspace*{-0.5 cm}

\begin{thebibliography}{}

\bibitem[Asplund \& Garcia Perez~(2001)]{agp2001} Asplund M., Garcia Perez A.E., 2001, A\&A, 372, 601

\bibitem[Behara et al.~(2009)]{behara2009} Behara N.T., et al., 2009, in preparation

\bibitem[Freytag et al.~(2002)]{fsd2002} Freytag B., Steffen M., Dorch B., 2002, AN, 323, 213

\bibitem[Kurucz~(1993)]{k1993} Kurucz, R.L., 1993, CD-ROM 13, SAO, http://cfaku5.cfa.havard.edu/

\bibitem[Sivarani et al.~(2006)]{siv2006} Sivarani T., Beers T.C., Bonifacio P., et al., 2006, A\&A, 459, 125

\bibitem[Wedemeyer et al.~(2004)]{wed2004} Wedemeyer~S., Freytag~B., Steffen~M., Ludwig~H.-G., Holweger~H.,
2004, A\&A\,414, 1121

\end{thebibliography}
\end{document}